\documentclass[a4paper]{jpconf}
\usepackage{graphicx}
\begin{document}
\title{Scholarly literature and the press: scientific impact and social perception of physics computing }

\author{ M. G. Pia$^{1}$, T. Basaglia$^{2}$, Z. W. Bell$^{3}$, P. V. Dressendorfer$^{4}$}

\address {$^1$INFN Sezione di Genova, 16146 Genova, Italy}
\address{$^2$CERN, 1211 Geneva, Switzerland}
\address{$^3$ORNL, Oak Ridge, TN 37830, USA}
\address{$^3$IEEE, Piscataway, NJ 08854, USA}

\ead{Maria.Grazia.Pia@cern.ch}
\begin{abstract}
The broad coverage of the search for the Higgs boson in the mainstream media is
a relative novelty for high energy physics (HEP) research, whose achievements
have traditionally been limited to scholarly literature. This paper illustrates
the results of a scientometric analysis of HEP computing in scientific
literature, institutional media and the press, and a comparative overview of
similar metrics concerning representative particle physics measurements. 
The picture emerging from these scientometric data documents the relationship
between the scientific impact and the social perception of HEP physics research
versus that of HEP computing. The results of this analysis suggest that improved
communication of the scientific and social role of HEP computing via press
releases from the major HEP laboratories would be beneficial to the high energy
physics community.

\end{abstract}

\section{Introduction}
The scientometric investigation summarized in this paper paper provides some
insights on the role of computing for High Energy Physics (HEP) both in the
context of scholarly publication and in the press.
It aims at assessing the potential contribution of this domain to communicate
the impact of high energy physics research on the society at large.

To carry out this analysis, a few landmark discoveries in particle physics have
been investigated, to determine the impact of such events on scientific
literature on one side, and their presence and visibility in the press on the
other side
in order to assess the social perception of this research domain.

The data sources used in this analysis were Thomson-Reuters' ISI Web of
Knowledge \cite{wos} database (covering 1970- to date), the collection of press
releases of representative high energy physics laboratories (CERN, Fermilab and
SLAC) and the online archives of some newspapers with strong circulation, such
as The New York Times, The Times, The Telegraph, the Frankfurter Allegemeine
Zeitung, Die Welt, Le Monde, Le Figaro, La Repubblica and Corriere della Sera.

The data collection for this analysis had to face some practical obstacles: the
publicly accessible archives of press releases of major HEP laboratories cover a
limited time span and provide limited search facilities; the online archives of
most newspapers offer only limited free access and permit in general simple
searches only.
These limitations hinder the feasibility of a rigorous statistical study;
nevertheless a consistent picture of the field emerges
from this investigation.

In the appraisal of the results reported in the following sections one should
take into account that 2013 statistics cover only approximately 2/3 of the year,
as they are limited to data collected before the CHEP 2013 conference.

\section{Citation patterns of high energy physics discoveries}

\begin{figure}[h]
\begin{center}
\begin{minipage}{14pc}
\includegraphics[width=14pc]{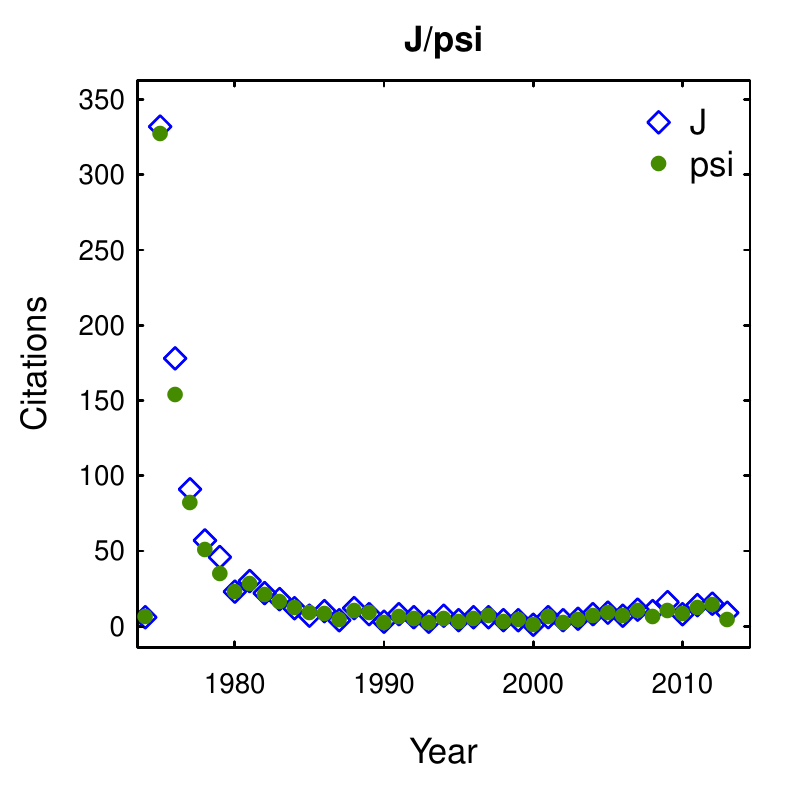}
\caption{\label{fig:Jpsi}Time profile of the citations of the papers documenting the discovery of the J/$\psi$.}
\end{minipage}\hspace{2pc}%
\begin{minipage}{14pc}
\includegraphics[width=14pc]{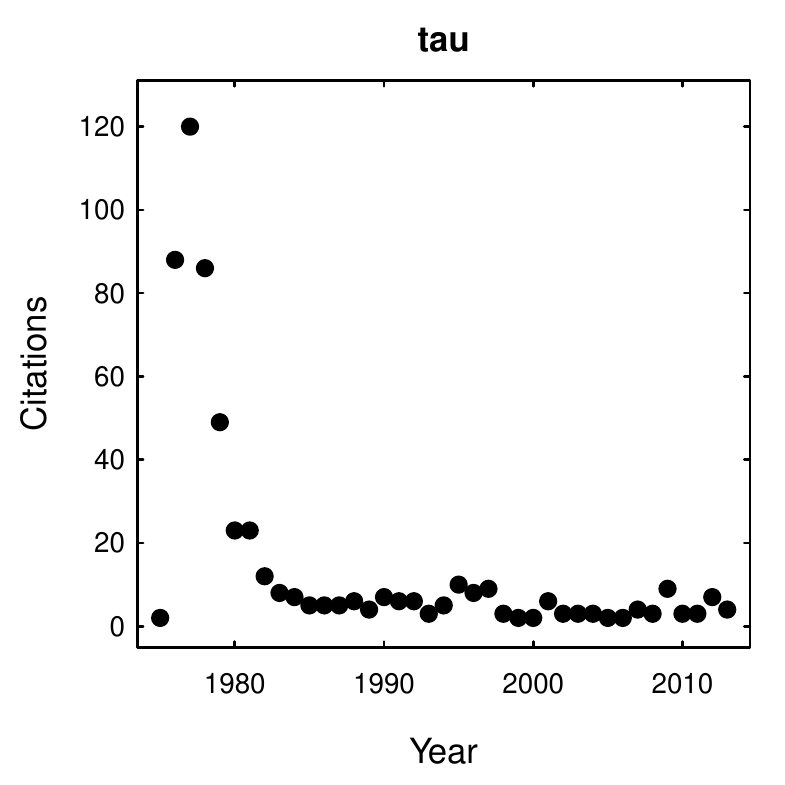}
\caption{\label{fig:tau}Time profile of the citations of the papers documenting the first observation of the $\tau$.}
\end{minipage} 
\end{center}
\end{figure}

\begin{figure}[h]
\begin{center}
\begin{minipage}{14pc}
\includegraphics[width=14pc]{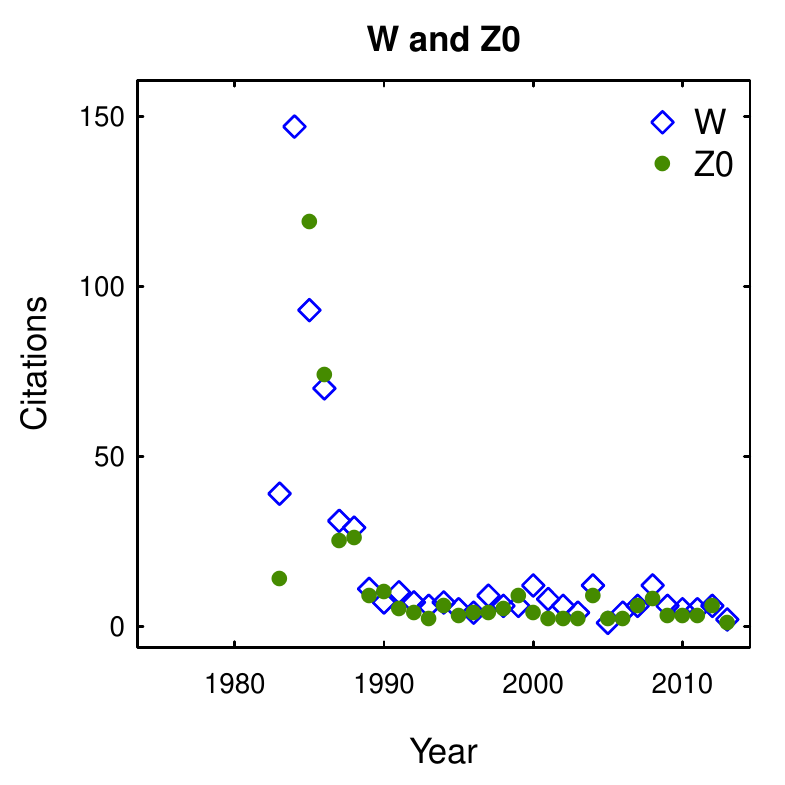}
\caption{\label{fig:wz}Time profile of the citations of the papers documenting the first observation of the W and Z$_0$ bosons.}
\end{minipage}\hspace{2pc}%
\begin{minipage}{14pc}
\includegraphics[width=14pc]{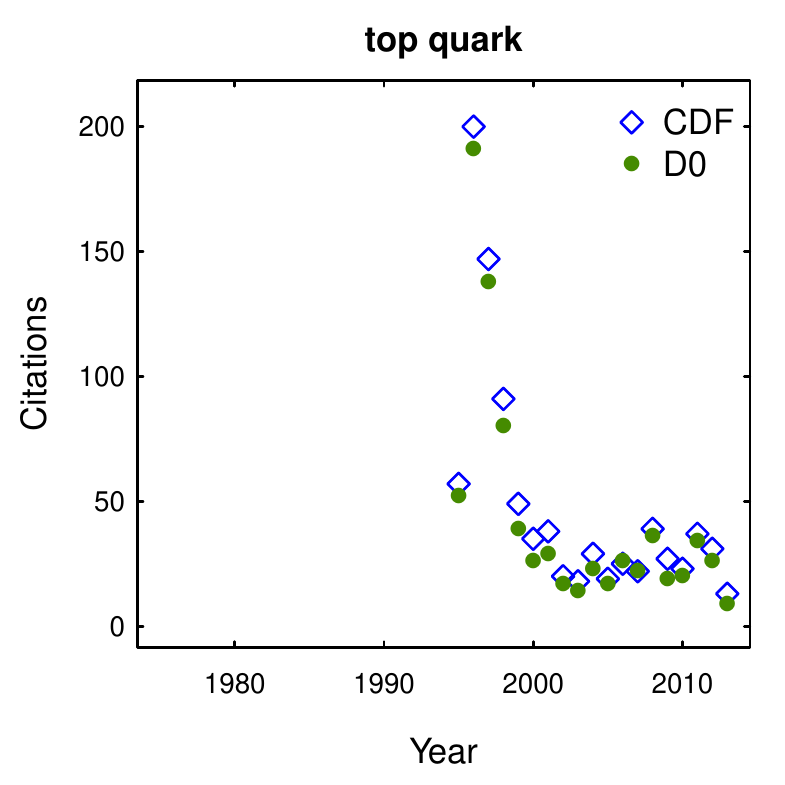}
\caption{\label{fig:topl}Time profile of the citations of the papers documenting the first observation of the top quark.}
\end{minipage} 
\end{center}
\end{figure}

The citation patterns of several papers associated with landmark
experimental discoveries have been investigated. 
They include the first
observation of the J/$\psi$ \cite{aubert,augustin}, the $\tau$ \cite{perl}, the
W and Z$_0$ \cite{arnison1, arnison2} at the CERN SPS collider, the top quark at
the Tevatron \cite{abe, abachi}, of neutrino oscillations \cite{fukuda}, and of
CP violation in the K and B meson systems \cite{alavi, na48, abe2, aubert2}.
The number of citations received by these papers is listed in Table \ref{tab:hepcite}.
The time evolution of the citations of these papers shows evident similarities:
the citation count exhibits an initial high peak, but it drops dramatically
after 3 to 5 years following the publication of the papers announcing the
discoveries.
Some examples are illustrated in Figures \ref{fig:Jpsi}-\ref{fig:topl}.

At this stage it is too early to evaluate whether the papers documenting the first 
observation of the Higgs boson will experience a similar citation rate.


\begin{table}
\caption{\label{tab:hepcite}Citations collected by representative HEP physics and software papers.}
\begin{center}
\begin{tabular}{| l |r|}
\hline
Subject &Citations \\
\hline
J/$\psi$ discovery &  1059, 963\\
$\tau$ discovery & 574 \\
W observation  & 590 \\
Z$_0$ observation & 550 \\
top quark observation & 1005 (CDF), 888  (D0)\\
neutrino oscillations & 2906 \\
CP violation (E832) & 346 \\
CP violation (NA48) &  262\\
CP violation (BaBar) & 335 \\
CP violation (Belle)  & 351 \\
Higgs boson observation & 730 (ATLAS), 695 (CMS)\\
MINUIT & 1330 \\
Geant4  & 4295 \\

\hline
\end{tabular}
\end{center}
\end{table}

\begin{figure}[h]
\begin{center}
\begin{minipage}{14pc}
\includegraphics[width=14pc]{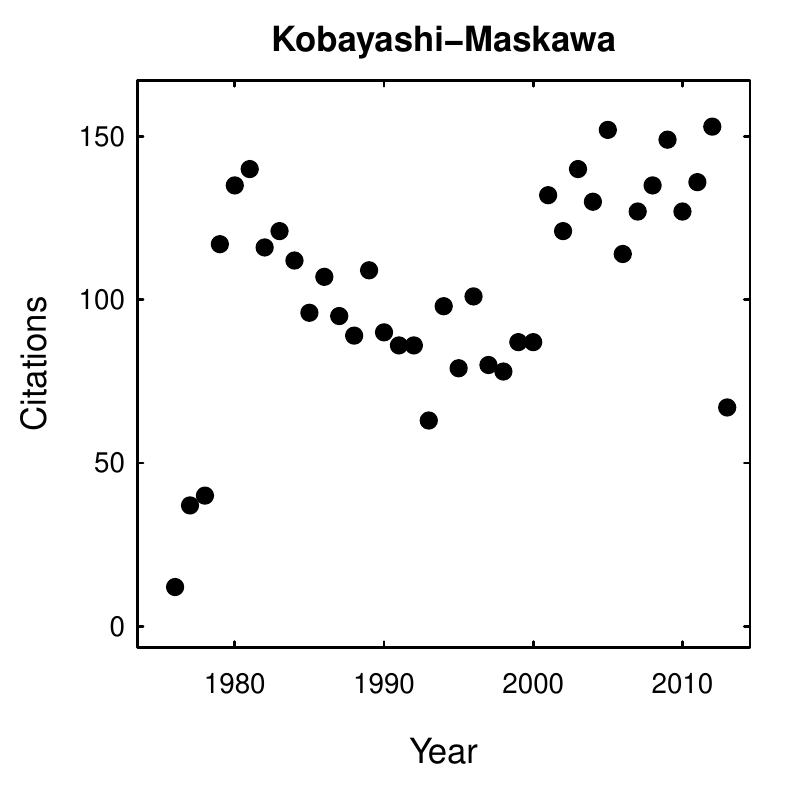}
\caption{\label{fig:KM}Time profile of the citations of the landmark Kobayashi-Maskawa paper.}
\end{minipage}\hspace{2pc}%
\begin{minipage}{14pc}
\includegraphics[width=14pc]{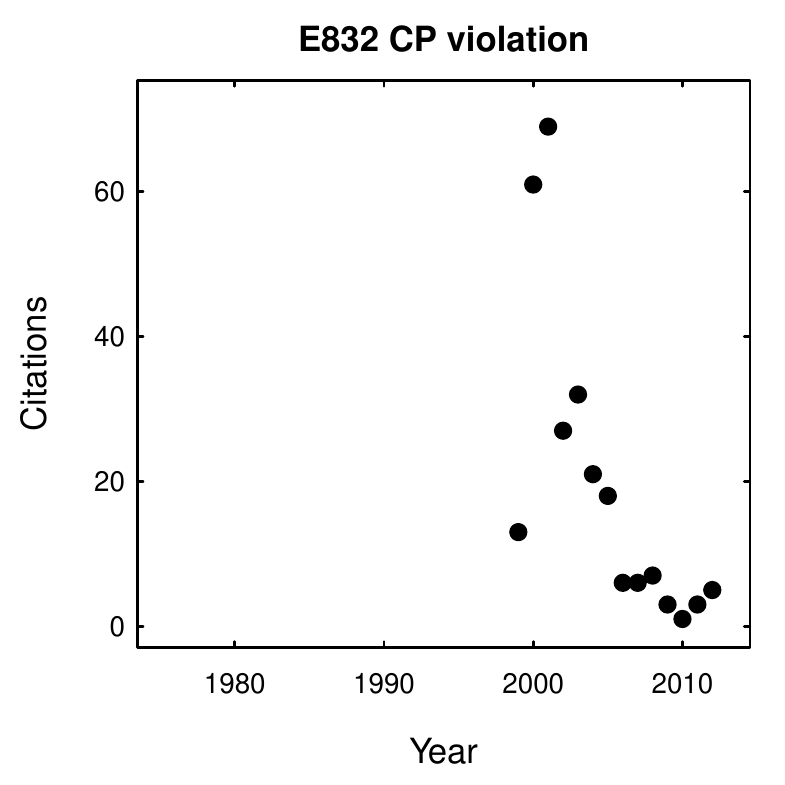}
\caption{\label{fig:E832}Time profile of the citations of the observation of CP violation in the E832 experiment.}
\end{minipage} 
\begin{minipage}{14pc}
\includegraphics[width=14pc]{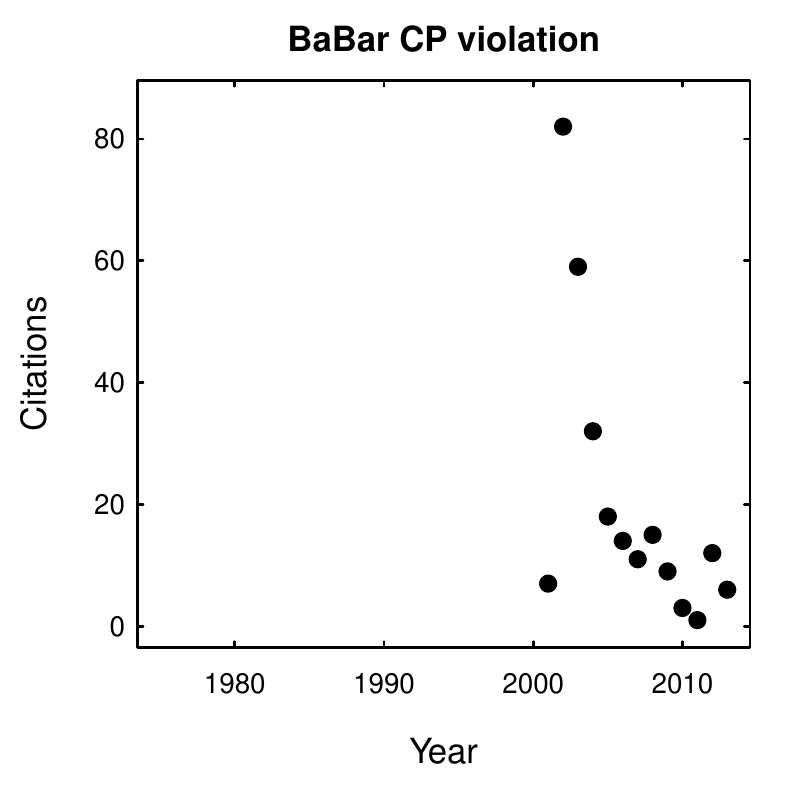}
\caption{\label{fig:babarcp}Time profile of the citations of the observation of CP violation in the BaBar experiment.}
\end{minipage} \hspace{2pc}%
\begin{minipage}{14pc}
\includegraphics[width=14pc]{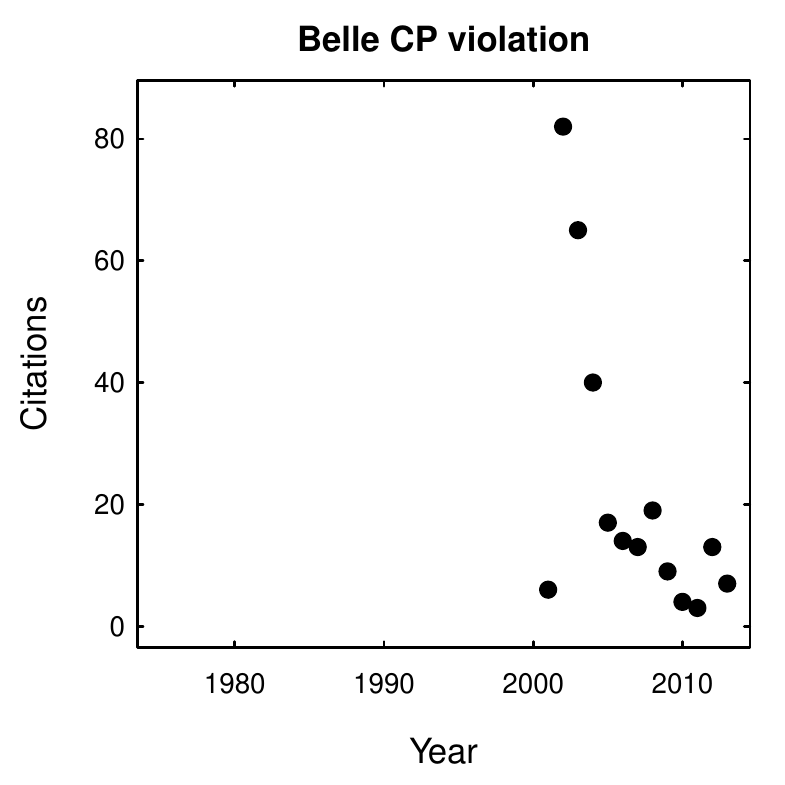}
\caption{\label{fig:belle}Time profile of the citations of the observation of CP violation in the Belle experiment.}
\end{minipage} 
\end{center}
\end{figure}

The relatively short citation lifetime of papers documenting major HEP experimental discoveries
contrasts with the extended citation life of fundamental theoretical papers in the same domain:
and example is shown in figures \ref{fig:KM}-\ref{fig:belle}, which report the citation profiles of the landmark
Kobayashi-Maskawa paper \cite{KM} and of representative experimental observations of CP violation
\cite{alavi, na48, abe2, aubert2}.

\section{Building awareness of HEP research in mainstream media}

When it comes to analyzing the impact in terms of visibility of experimental
particle physics discoveries in the press, one can observe a correlation between
their presence in mass-circulation newspapers and press releases issued by major
particle physics laboratories.

The high visibility in the mainstream media of the discovery of the Higgs boson
is well known: admittedly, it was the result of a deliberate communication
effort to promote particle physics.
We found a number of other cases, where formal communication to the press of
events related to high energy physics leads to articles in widely read
newspapers.

As an example, one can take the announcement about the observation of
matter-antimatter asymmetry in the LHCb experiment (CERN press release of the
24th of April 2013), which is reflected on the same day in articles published on
The Telegraph and the Italian Corriere della Sera.
Other examples are the announcements of the production of antihydrogen and a new
state of matter (CERN press releases of the 18th September 2002 and the 10th
February 2000, respectively), which which raised immediate and long-lasting
attention in the press: articles on this subject were published on various
newspapers in the next days, such as The New York Times, Le Monde, Le Figaro, La
Repubblica, Die Welt and the Frankfurter Allgemeine Zeitung.
The scholarly publication \cite{athena} documenting the production of
antihydrogen had received 463 citations at the time of the CHEP 2013 conference.

Similarly, the discovery of the top quark announced by Fermilab in a press
release on the 2nd March 1995 found immediate echo in the press both in US and
European newspapers.

Correlation between press releases by major laboratories and articles in
mass-circulation newspapers is observed not only regarding major particle physics
discoveries, but also technological achievements: examples are the the
lowering of the heaviest part of the CMS detector in the LHC tunnel (CERN press
release of the 28th February 2007) and an internet speed record achieved in
transferring data between CERN and Caltech (CERN press release of 15th October
2003): the corresponding stories can be found in newspapers such as Die Welt, La
Repubblica and The New York Times in the days following the press release.

These examples (and others not mentioned here due to the limited page allocation in these
conference proceedings) show that the press releases are an effective strategy for communicating 
the scientific advances to the general public.



\section{HEP computing and scientific software}

If we now move to the computing domain, we observe that only the Grid reaches
high and long-lasting visibility in the mainstream press, besides the Web
technology development.
Stories in major newspapers related to these subjects are associated with a
significant number of CERN press releases.

If we now turn our attention to papers on software development related to
experimental particle physics, we note that some of them obtain a high score in
terms of citations, but also enjoy a long-lasting impact in the scientific
publishing arena.
Representative cases in this domain are MINUIT \cite{minuit} and Geant4 \cite{g4nim,g4tns}.

\begin{figure}[h]
\begin{center}
\begin{minipage}{16pc}
\includegraphics[width=16pc]{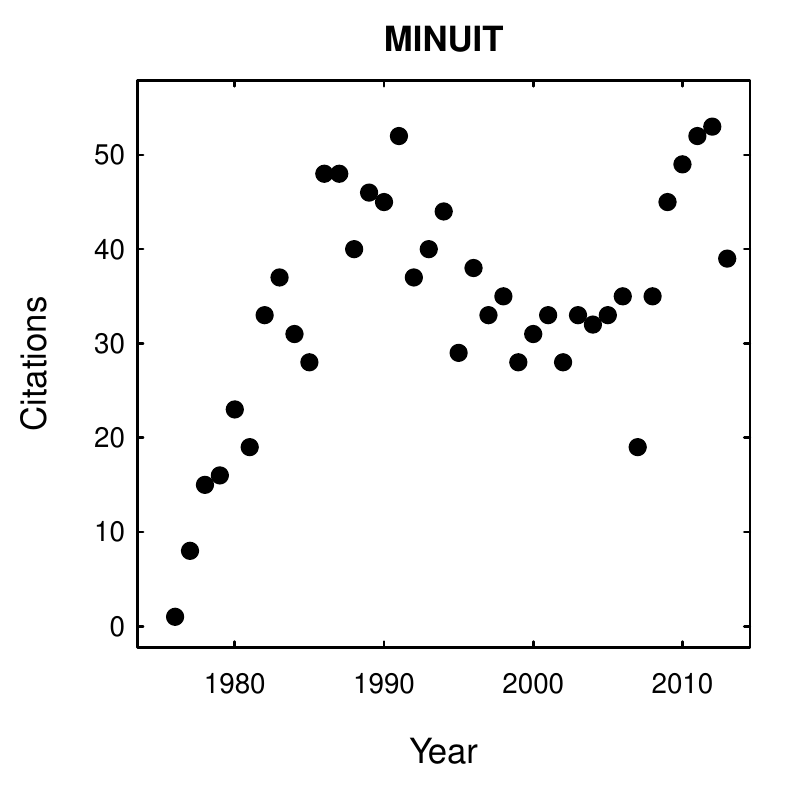}
\caption{\label{fig:minuit}Time profile of the citations of MINUIT.}
\end{minipage}\hspace{2pc}%
\begin{minipage}{14pc}
\includegraphics[width=14pc]{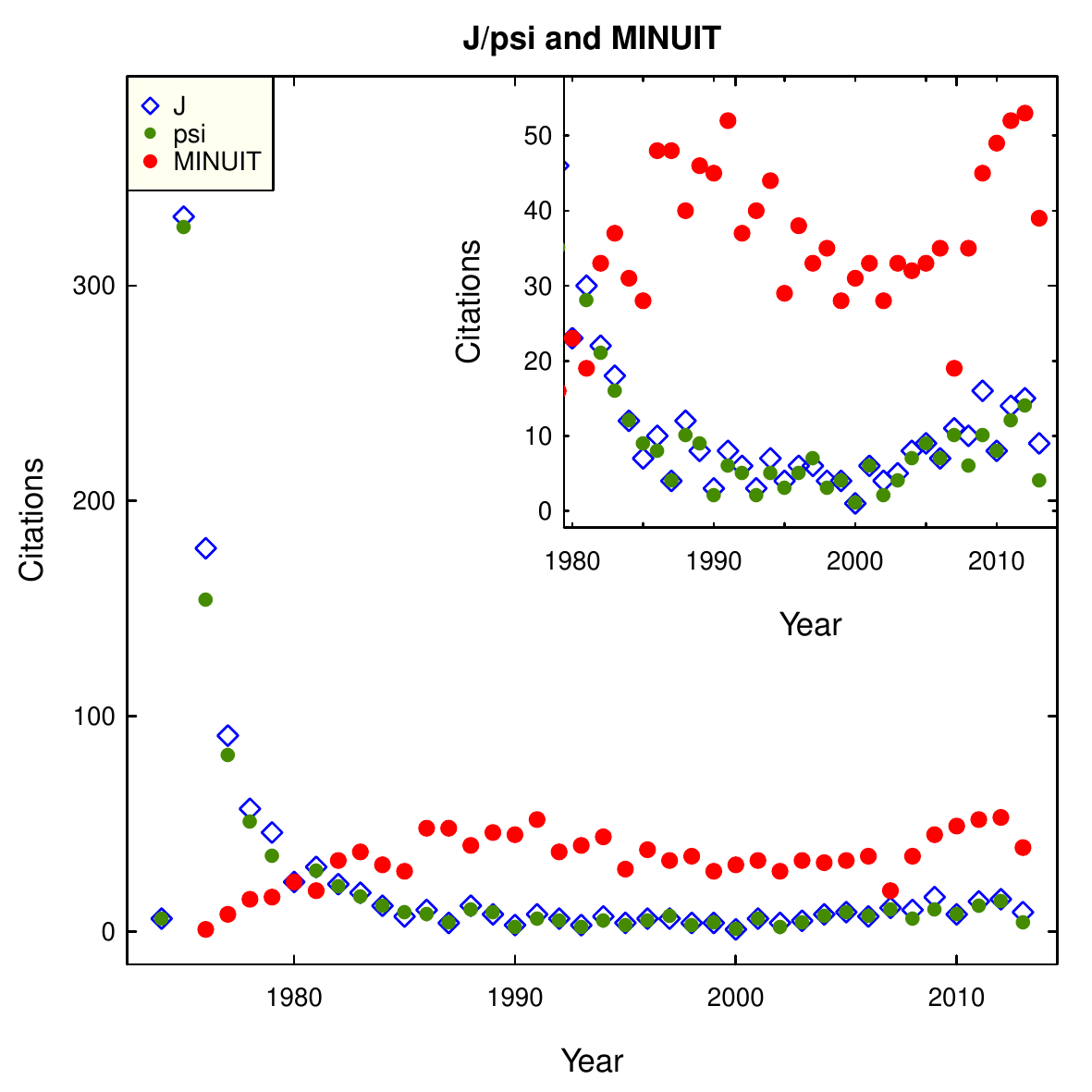}
\caption{\label{fig:Jpsi_minuit}The time profile of the citations of the MINUIT paper and the almost contemporary papers about the
discovery of the J/$\psi$.}
\end{minipage} 
\end{center}
\end{figure}

MINUIT reference paper \cite{minuit} was published shortly after the discovery
of the J/$\psi$; at the time of the CHEP 2013 conference it had received 1330
citations, distributed as shown in Figure \ref{fig:minuit}.
It is worthwhile to note in Figure \ref{fig:Jpsi_minuit} that approximately five
years after the discovery of the J/$\psi$ the number of MINUIT yearly citations
was comparable to that of this landmark physics discovery, while nowadays the
citations of MINUIT largely outnumber those received by the J/$\psi$ papers.


The first Geant4 reference paper \cite{g4nim} had received 4295 citations at
the time of the CHEP 2013 conference.
It is the most cited paper in the categories of Particle and Fields Physics
(encompassing 264075 papers), Nuclear Science and Technology, and Instruments
and Instrumentation (jointly encompassing 618147 papers).
These numbers result from a citation analysis performed on 9 October 2013, based
on Thomson-Reuters' Web of Science \cite{wos}.
This paper is also the most cited publication of CERN (i.e. including CERN authors).

To the best of our efforts, no mention of HEP scientific software and its achievements
could be found in CERN press releases, nor in the mainstream press.
Given the observed correlation between visibility to the general public and communication
from major laboratories, one could extrapolate a similar correlation also regarding the
lack of attention to this HEP research domain.

\begin{figure}
\centerline{\includegraphics[angle=0,width=16cm]{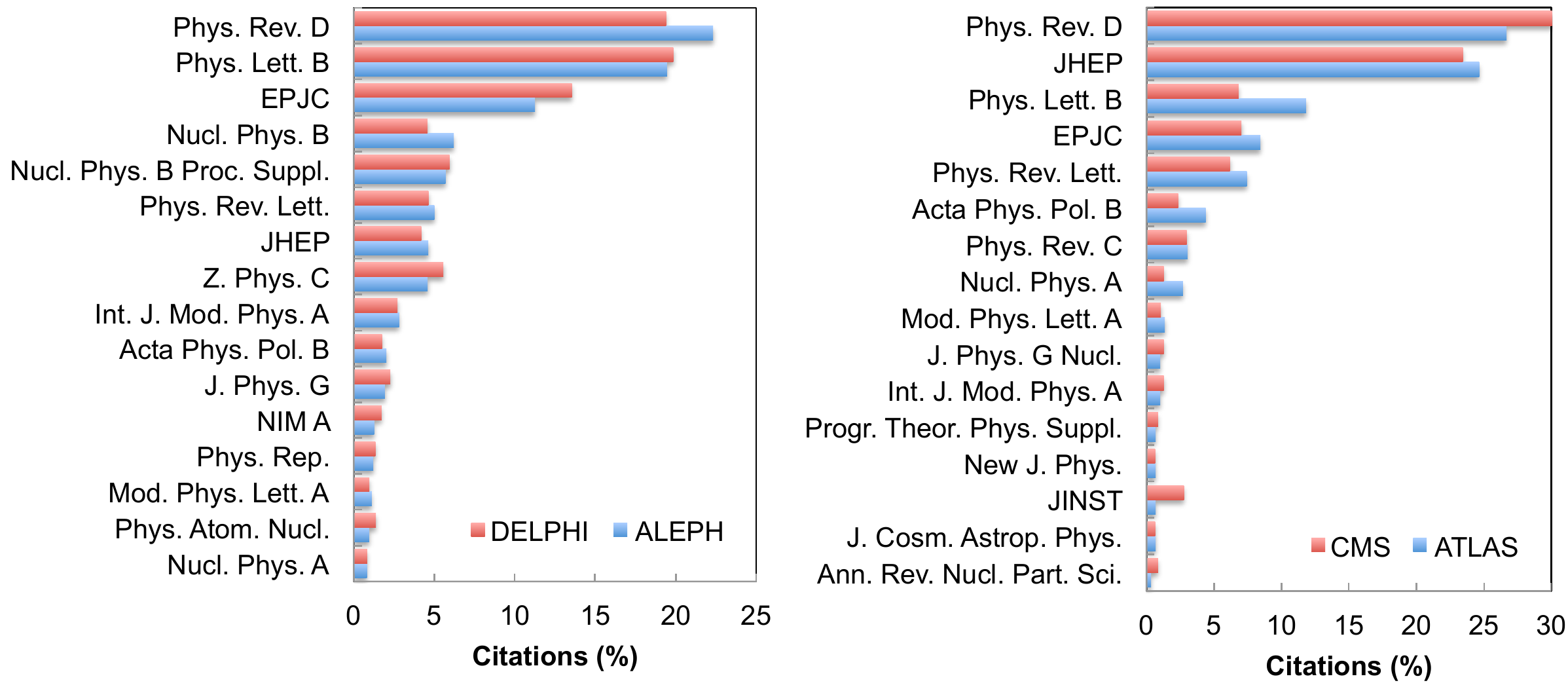}}
\caption{Sources of citations of physics papers published by representative LEP
experiments (ALEPH and DELPHI, on the left) and LHC experiments (ATLAS and CMS, on the right); the histograms include more than
90\% of the citations received by the physics papers of the selected
experiments.}
\label{fig_exp_physcite}
\end{figure}

The citations of these HEP software systems exhibit a distinctive pattern regarding their origin,
which is largely different from the typical patterns associated with HEP papers.
The vast majority of the citations of experimental HEP publications
originate from the same subject area and a few closely related fields, such as
astroparticle physics and nuclear physics:
as an example, the journals from which more than 90\% of the citations of representative LEP and LHC physics papers originate 
are shown in Figure \ref{fig_exp_physcite}.

\begin{figure}[h]
\begin{center}
\begin{minipage}{14pc}
\includegraphics[width=14pc]{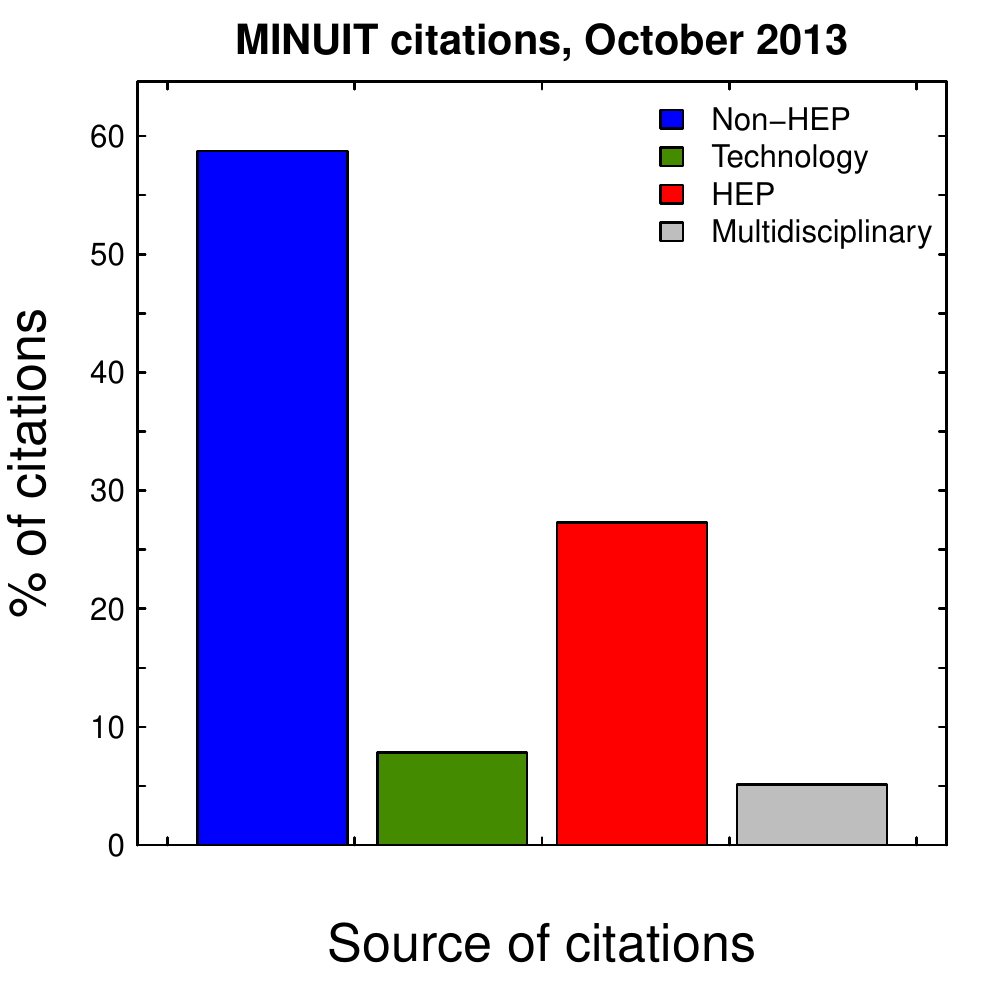}
\caption{\label{fig:minuitorigin}Domains of origin of the MINUIT citations.}
\end{minipage}\hspace{2pc}%
\begin{minipage}{14pc}
\includegraphics[width=14pc]{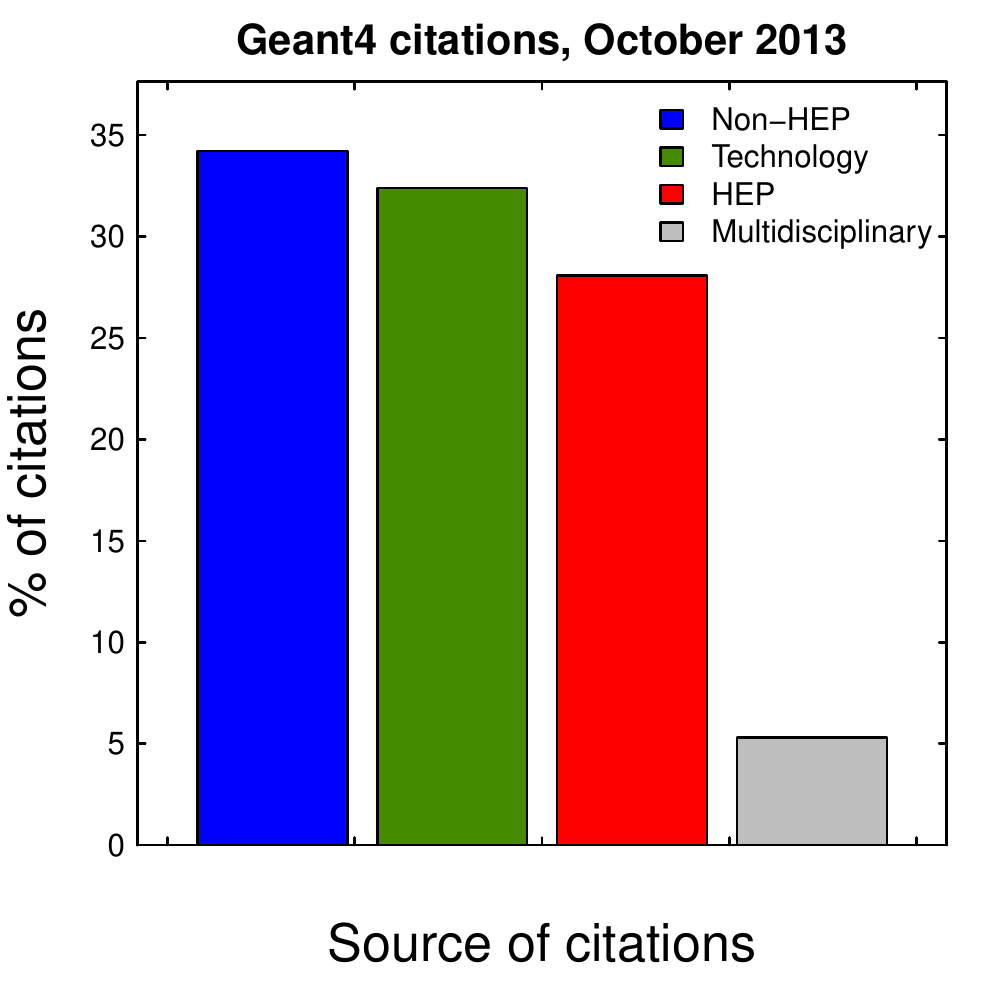}
\caption{\label{fig:g4origin}Domains of origin of Geant4 citations.}
\end{minipage} 
\end{center}
\end{figure}

The citations of MINUIT and Geant4 show a largely multidisciplinary character,
as is illustrated in Figures \ref{fig:minuitorigin} and \ref{fig:g4origin}.
For both software systems the citations originating from areas other than high energy physics outnumber those deriving from HEP.
These citation patterns document the impact of the technological research motivated by HEP
experiments on other domains, some of which have evident social impact (e.g. the large number of Geant4
citations originating from the bio-medical environment).

\section{Conclusions}

This scientometric analysis has assessed some distinctive patterns in scholarly publications and visibility in
mainstream media of high energy physics experimental discoveries and scientific software systems.

Major experimental discoveries are characterized by short-lived citation patterns, which exhibit a rapid drop
shortly after the observation.
Major HEP software systems, such as MINUIT and Geant4, collect citations over an extended period, which 
demonstrates their continued contribution to the advancement of science.
Their citations have a multidisciplinary nature, which contrasts with the narrower scope of the citations 
collected by HEP experiments, mostly originating from the HEP community itself and a few closely related domains.

Visibility in mainstream media appears associated with communications to the press issued by major 
HEP institutes. 
Press releases and the associated stories in mass-circulation media contribute to determine the social 
perception of high energy physics research among the general public.

The results of this scientometric analysis document the impact of scientific software research motivated by HEP experiments
on domains other than high energy physics, including some of evident social relevance.
They could provide objective support for improved communication of the benefits of HEP research and developments
via press releases from the major HEP institutes.

\section*{Acknowledgments}

The authors wish to thank the CERN Library for providing essential tools to this scientometric analysis.


\section*{References}

\begin{thebibliography}{99}

\bibitem{wos}
http://apps.webofknowledge.com

\bibitem{aubert}
J. J. Aubert et al. 1974
\textit{ Phys. Rev. Lett.} {\bf 33} 1404

\bibitem{augustin} 
J. E. Augustin et al. 1974
\textit{ Phys. Rev. Lett.} {\bf 33} 1406


\bibitem{perl}
M. L. Perl 1975 
\textit{Phys. Rev. Lett.} {\bf 35} 1489

\bibitem{arnison1}
G. Arnison 1983
\textit{Phys. Lett. B } {\bf 122}  103

\bibitem{arnison2} 
G. Arnison 1983 
\textit{Phys. Lett. B} {\bf 126} 398

\bibitem{abe}
F. Abe et al. 1995
\textit{ Phys. Rev. Lett. } {\bf 74} 2626

\bibitem{abachi}
S. Abachi 1995
\textit{ Phys. Rev. Lett. } {\bf 74} 2632


\bibitem{fukuda}
Y. Fukuda et al. 1988
{\textit Phys. Rev. Lett.} {\bf 81} 1562


\bibitem{alavi}
A. Alavi-Harati et al. 1999
{\textit  Phys. Rev. Lett. } {\bf 83} 22

\bibitem{na48}
V. Fanti et al. 1999
{\textit  Phys. Lett. B} {\bf 465} 335 

\bibitem{abe2}
K. Abe et al. 2001 
{\textit  Phys. Rev. Lett.  } {\bf 87} 091802

\bibitem{aubert2}
B. Aubert et al.2001
{\textit   Phys. Rev. Lett.   } {\bf 87} 091801

\bibitem{KM}
M. Kobayashi and  T. Maskawa 1973
{\textit  Progr. Theor. Phys.   } {\bf } 652


\bibitem{athena}
M. Amoretti et al. 2002
\textit{Nature } {\bf  419} 456



\bibitem{minuit}
F. James and M. Roos 1975
\textit{Comp. Phys. Comm.} {\bf 10} 343


\bibitem{g4nim}
Agostinelli S \textit{et al} 2003
\textit{Nucl. Instrum. Meth. A} {\bf 506} 250

\bibitem{g4tns}
Allison J \textit{et al}  2006
\textit{IEEE Trans. Nucl. Sci.} {\bf 53} 270




\end{thebibliography}
\end{document}